# Topologically protected surface states in a centrosymmetric superconductor $\beta$-PdBi$_2$

M. Sakano[1], K. Okawa[2], M. Kanou[2], H. Sanjo[1], T. Okuda[3], T. Sasagawa[2], K. Ishizaka[1]

[1]*Quantum-Phase Electronics Center (QPEC) and Department of Applied Physics,*
*The University of Tokyo, Tokyo 113-8656, Japan*
[2]*Materials and Structures Laboratory, Tokyo Institute of Technology, Kanagawa 226-8503, Japan*
[3]*Hiroshima Synchrotron Radiation Center, Hiroshima University, Higashi-Hiroshima 739-0046, Japan*

The topological aspects of electrons in solids emerge in realistic matters as represented by topological insulators[1-3]. They are expected to show a variety of new magneto-electric phenomena, and especially the ones hosting superconductivity are strongly desired as the candidate for topological superconductors (TSC)[2-7]. Possible TSC materials have been mostly developed by introducing carriers into topological insulators[8-14], nevertheless, those exhibiting indisputable superconductivity free from inhomogeneity are very few[15]. Here we report on the observation of topologically-protected surface states in a centrosymmetric layered superconductor, $\beta$-PdBi$_2$, by utilizing spin- and angle-resolved photoemission spectroscopy. Besides the bulk bands, several surface bands, some of which crossing the Fermi level, are clearly observed with symmetrically allowed in-plane spin-polarizations. These surface states are precisely evaluated to be topological, based on the $Z_2$ invariant analysis in analogy to 3-dimensional strong topological insulators. $\beta$-PdBi$_2$ may offer a TSC realized without any carrier-doping or applying pressure, i.e. a solid stage to investigate the topological aspect in the superconducting condensate.

Topological insulators are characterized by the non-trivial $Z_2$ topological invariant acquired when the conduction and valence bands are inverted by spin-orbit interaction (SOI), and the gapless surface-state appears[1-3]. This topologically non-trivial surface state possesses the helical spin-polarization locked to momentum, and is expected to host various kinds of new magneto-electric phenomena. Especially, the ones realized with superconductivity are theoretically investigated as the candidate for TSC[2-4], whose excitation is described as Majorana Fermions, the hypothetical particles originating from the field of particle physics[5-7]. Experimentally, several superconductors developed by utilizing topological insulators are reported thus far, such as Cu-intercalated $Bi_2Se_3$[8-11], In-doped SnTe[12], and $M_2Te_3$ ($M$ = Bi, Sb) under pressure[13,14]. While the previous studies of point-contact spectroscopy on $Cu_xBi_2Se_3$[8, 9] and In-SnTe[12] suggest the existence of Andreev bound state thus raising the possibility of TSC, the scanning tunneling microscope / spectroscopy (STM/STS) reports the simple s-wave like full SC gap[16]. Theoretically, this contradiction has been discussed in terms of the possible peculiar bulk odd-parity pairing[17], which awaits the experimental verifications by various probes[18,19]. However, partly due to the inhomogeneity effect accompanied by doping or pressurizing, the unambiguous clarification of TSC in doped topological insulators has been hindered until now. Recently reported half Heusler superconductor RPtBi (R: rare earth) is another class of material as the candidate for TSC[15,20]. Practically, however, its low critical temperature ($T_c$ < 2 K) and the noncentrosymmetric crystal structure without a unique cleavage plane may be raising some difficulties on its further investigation.

In this Letter, we introduce $\beta$-PdBi$_2$ as a new candidate material of TSC. $\beta$-PdBi$_2$ is a superconductor with a centrosymmetric tetragonal crystal structure of space group I4/mmm[21-23] as shown in Fig. 1a, a much simpler one compared to the related noncentrosymmetric superconductor $\alpha$-PdBi recently being discussed as a possible TSC[24,25]. Pd atoms, each of them located at the center of the square-prism of 8 Bi atoms, form the layered body-centered unit cell. Pd-Bi$_2$ layers are stacking in Van der Waals nature, making it a feasible compound for cleaving. We investigate the electronic structure of $\beta$-PdBi$_2$ by using (spin-) angular- resolved photoemission spectroscopy, (S)ARPES. With the large single crystals of good quality, exhibiting the high residual resistivity ratio (~14) and a clear superconducting transition at $T_c = 5.3$ K (Fig. 1a), several spin-polarized surface-states are clearly observed in addition to the bulk bands. Based on the relativistic first-principles calculation on bulk and the slab calculation on surface, we find that the observed surface-states can be unambiguously interpreted to be topologically non-trivial.

The band structure of $\beta$-PdBi$_2$ observed by ARPES is shown in Fig. 1c,d. For simply describing the (S)ARPES results hereafter, we use the projected 2-dimensional (2D) surface Brillouin zone (BZ) depicted in Fig. 1b by a green square. The projected high-symmetry points are $\bar{\Gamma}$, $\bar{M}$ and $\bar{X}$, and we define $k_x$ as the momentum along $\bar{\Gamma}$-$\bar{M}$. The ARPES image in Fig. 1d is recorded along $\bar{X}$-$\bar{\Gamma}$ (left) and $\bar{\Gamma}$-$\bar{M}$ (right), respectively. Bands crossing the Fermi level ($E_F$) are predominantly derived from Bi 6$p$ components with large dispersions from the binding energy ($E_B$) of $E_B \sim 6$ eV to above $E_F$. On the other hand, bands mainly consisting of Pd 4$d$ orbitals are located around $E_B = 2.5 \sim 5$ eV with rather small dispersions. Near $E_F$, two hole bands ($\alpha$, $\beta$) and one electron band ($\gamma$) are observed along $\bar{\Gamma}$-$\bar{M}$, whereas for $\bar{X}$-$\bar{\Gamma}$, the large ARPES intensity from another electron band ($\delta$) is additionally observed. As we can see in Fig. 1c, the experimental Fermi surface (FS) mapping mostly well agrees with the 2D projection of the calculated bulk FS (Fig. 1b).

To compare with ARPES, the calculation of bulk band dispersions projected into 2D BZ is shown in Fig. 1e. Considering that the ARPES intensity includes the integration of finite $k_z$-dispersions due to the surface sensitivity, the overall electronic structure is in a good agreement with the calculation, nevertheless, several differences can be noticed. The most prominent one appears in the orange rectangles in Fig. 1d,e. A sharp Dirac-cone-like dispersion is experimentally observed where the calculated bulk bands show a gap of ~0.55 eV around $\bar{\Gamma}$-point. To confirm its origin, we performed a slab calculation for 11 PdBi$_2$ layers (Fig. 1f). Apparently, a Dirac-cone-type dispersion appears in the gapped bulk states, showing a striking similarity to ARPES (Fig. 1d). It clearly presents the surface origin of this Dirac-cone band.

Now we focus on the observed surface Dirac-cone (S-DC) band. The close-up of the S-DC is demonstrated in Fig. 2a, indicating its crossing point at $E_B = E_D = 2.41$ eV ($E_D$: the energy of Dirac-point). Such a clear Dirac-cone shaped band strongly reminds us of the helical edge states in 3D strong topological insulators (STI). We can see the very isotropic character of S-DC in its constant-energy cuts (Fig. 2b), appearing as the perfectly circular-shaped contour even at $E - E_D = 0.8$ eV with a large momentum radius of 0.3 Å$^{-1}$. It is in contrast to the warping effect often appearing in trigonal STIs[26,27]. The spin polarization of S-DC is also directly confirmed by SARPES experiments as depicted in Fig. 2c[28]. Figure 2e,f show the results for $y$-component spin, measured along $k_x$ ($\overline{\Gamma}$-$\overline{M}$). Because of $C_{4v}$ symmetry, $x$- and $z$-components are forbidden (see 1.3 in SI). The red (blue) curves in Fig. 2f, indicating the energy distribution curves (EDCs) of spin-up (-down) components, clearly show the spin-polarized band dispersions. As easily seen in the SARPES image (Fig. 2e), the spin polarization with spin-up (spin-down) for negative (positive) dispersion of S-DC is confirmed. The observed spin-polarized S-DC thus presents a strong resemblance to the helical surface state in STI.

To evaluate whether the observed surface-state is topologically non-trivial, we derive $Z_2$ invariant $v_0$ for $\beta$-PdBi$_2$, in analogy to 3D-STI[29]. For 3D band insulators with inversion symmetry, $v_0$ obtained from the parity eigenvalues of filled valence bands at eight time-reversal invariant momenta (TRIM) classifies whether it is STI ($v_0 = 1$) or not ($v_0 = 0$). The bulk $\beta$-PdBi$_2$ is apparently a metal, nevertheless, here we define a "gap" in which there is no crossing of the bulk band dispersions through the entire BZ. By considering the "gap", we discuss its topological aspect by calculating $v_0$. The calculated bulk bands with and without SOI are shown in Fig. 3a,b, respectively. The valence bands are identified by numbering as indicated on the right side of respective graphs. Note that all bands are doubly spin-degenerate. By comparing Fig. 3a,b, we notice that many anticrossings are introduced by SOI, including the ~0.55 eV gap opening in the green rectangle region where S-DC appears. Here we focus on "gap7-6" between bulk bands 7 and 6 (B7 and B6), shaded by pink in Fig. 3b (see 2.1 in SI for "gap" assessment). In analogy to 3D-STI, we derive $v_0$ from the parity indices of bands below "gap7-6". Due to I4/mmm symmetry, $v_0$ can be calculated by considering solely Γ and Z points, whose symmetries of wave-functions are listed in Fig. 3d for respective bands (see 2.2 in SI for $v_0$ analysis). Those indicated by red (black) is of odd (even) parity. The $Z_2$ invariant $v_0$ can be thus obtained for respective "gap" as indicated in Fig. 3d. We find that "gap 7-6" is characterized by $v_0 = 1$, indicating its analogy to 3D-STI. This requires an odd number of surface-states in $\overline{\Gamma}$-$\overline{M}$, connecting B7 and B6, to topologically link the bulk $β$-PdBi$_2$ and a vacuum. The observation of spin-helical S-DC in "gap 7-6" clearly represents the characters of such topologically-protected surface-state.

By further looking at the list of $v_0$ in Fig. 3d, we notice $v_0 = 1$ for "gap 9-8" shaded by blue in Fig. 3b (see 2.1 in SI). It clearly indicates that topological surface-states connecting B9 and B8 must exist, where we may observe the effect of superconductivity if located close enough to $E_F$. To clarify this possibility, the close-up of ARPES image near $E_F$ is shown with the calculation in Fig. 4a,b. The green curves in Fig. 4a (bottom) indicate the calculated surface-states crossing $E_F$ separately from the 2D projected bulk bands shaded by gray. They appear at the smaller-$k_x$ side of $β$(B8) and $γ$(B9) bands. Experimentally, the sharp peaks indicative of 2D surface-states are observed in momentum-distribution-curve (MDC) at $E_F$, as denoted by S1 and S2 in Fig. 4a (top). As shown in the calculated band image in Fig. 4e, S2 should be the topological surface-state connecting B9 and B8, whereas S1 appearing in "gap 8-7" must be trivial (see 2.3 in SI for detailed description of S2).

The spin polarization of topological S2 as well as trivial S1 is also confirmed experimentally. As shown in Fig. 4b, the y-oriented spin polarizations of S1 (#2-5) and S2 (#7-10) along $k_x$ ($\overline{\Gamma}$-$\overline{M}$) are clearly observed in the spin-resolved EDC. Here, the peak positions for S1 / S2 (bulk β) bands are depicted by green circles (black squares). We can see that S1 and S2 are both spin-polarized with spin-up for $k_x > 0$, whereas they get inverted for $k_x < 0$ (Fig. 4c,d) as required by the time-reversal symmetry. These clearly indicate that both topological and trivial surface-states crossing $E_F$ possess the in-plane spin-polarizations.

Here we note that spin-polarized topological S2 and S-DC are both derived as a consequence of SOI, but in different processes. For the case of S2 in "gap 9-8", we see that $v_0$ changes to 1 by including SOI, thus indicating the SOI-induced band inversion of B8-B10 occurring at $\Gamma$ (see Fig. 3c,d). This situation is fairly similar to the topological phase transition being discussed in 3D-STI[30]. For S-DC in "gap 7-6", on the other hand, $v_0 = 1$ is realized already in the non-relativistic case (Fig. 3c), due to the inversion of $A_{1g}$ and $A_{2u}$ bands introduced by Bi$6p$ - Pd$4d$ mixing. This non-relativistic situation should be rather analogous to the 3D Dirac semimetals[31,32] as represented by the bulk Dirac-points appearing along Z-M and Z-X (Fig. 3a), which may accompany the spin-degenerate surface-states (Fermi arcs). The role of SOI in this case is the gap opening at these bulk Dirac-points, giving rise to the spin-polarized S-DC connecting the gap-edges.

The next future step for $\beta$-PdBi$_2$ should be the direct elucidation of the superconducting state. Low temperature ultrahigh-resolution ARPES will surely be a strong candidate for such investigation[33,34]. There may be a chance to observe non-trivial superconducting excitations, by selectively focusing on the surface and bulk band dispersions as experimentally presented in Bi$_2$Se$_3$ / NbSe$_2$ thin film[34]. STM/STS, on the other hand, can locally probe the SC state around the vortex cores. As theoretically suggested, it may capture the direct evidence of Majorana mode[4,10,35,36]. We should note that $\beta$-PdBi$_2$ will also provide a solid platform for bulk measurements such as thermal conductivity and NMR, which are expected to give some information on the odd-parity superconductivity[18,19]. It may thus contribute to making the realm of superconducting topological materials, and pave the way to various new findings such as the direct observation of Majorana fermions dispersion and/or surface Andreev bound states[35,36], clarification of its relation to the possible odd-parity superconductivity[10,17], bulk-surface mixing effect[35,37], and so on.

**Methods**

Single Crystals of $\beta$-PdBi$_2$ were grown by a melt growth method. Pd and Bi at a molar ratio of 1:2 were sealed in an evacuated quartz tube, pre-reacted at high temperature until it completely melted and mixed. Then, it was again heated up to 900°C, kept for 20 h, cooled down at a rate of 3°C/h down to 500°C, and rapidly quenched into cold water. The obtained single crystals had good cleavage, producing flat surfaces as large as ~ 1 × 1 cm$^2$.

ARPES measurement ($h\nu$ = 21.2) were made at the Department of Applied Physics, The University of Tokyo, using a VUV5000 He-discharge lamp and an R4000 hemispherical electron analyzer (VG-Scienta). The total energy resolution was set to 10 meV. Samples were cleaved *in situ* at around room temperature and measured at 20 K.

SARPES with the HeI$\alpha$ light source ($h\nu$ = 21.2 eV) was performed at the Efficient SPin REsolved SpectroScOpy (ESPRESSO) end station attached to the APPLE-II type variable polarization undulator beam line (BL-9B) at the Hiroshima Synchrotron Radiation Center (HSRC)[22]. The analyzer of this system consists of 2 sets of VLEED (Very low energy electron diffraction) spin detectors, thus enabling the detection of the electron spin orientation in 3 dimension. The angular resolution was set to ±1.5° and the total energy resolution was set to 35 meV. Samples were cleaved *in situ* at around room temperature and measured at 20 K.

**Acknowledgements**

We thank R. Arita for fruitful discussion, A. Kimura, H. Namatame, M. Taniguchi for sharing SARPES infrastructure. M.S. is supported by Advanced Leading Graduate Course for Photon Science (ALPS). M.S. and M.K. are supported by a research fellowship for young scientists from JSPS. This research was partly supported by Precursory Research for Embryonic Science and Technology (PRESTO), Japan Science and Technology Agency; the Photon Frontier Network Program of the MEXT; Research Hub for Advanced Nano Characterization, The University of Tokyo, supported by MEXT, Japan; and Grant-in-Aid for Scientific Research from JSPS, Japan (KAKENHI Kiban-B 24340078 and Kiban-A 23244066).

**Author contributions**

M.S. and H. S. carried out ARPES. M. S., T. O and K. I. carried out SARPES. K.O, M. K and T. S. synthesized and characterized the single crystals. T. S. carried out the calculations. M. S. and K. I. analysed (S)ARPES data and wrote the manuscript with input from K. O., M. K., T. O. and T.S. K.I. conceived the experiment. All authors contributed to the scientific discussions.

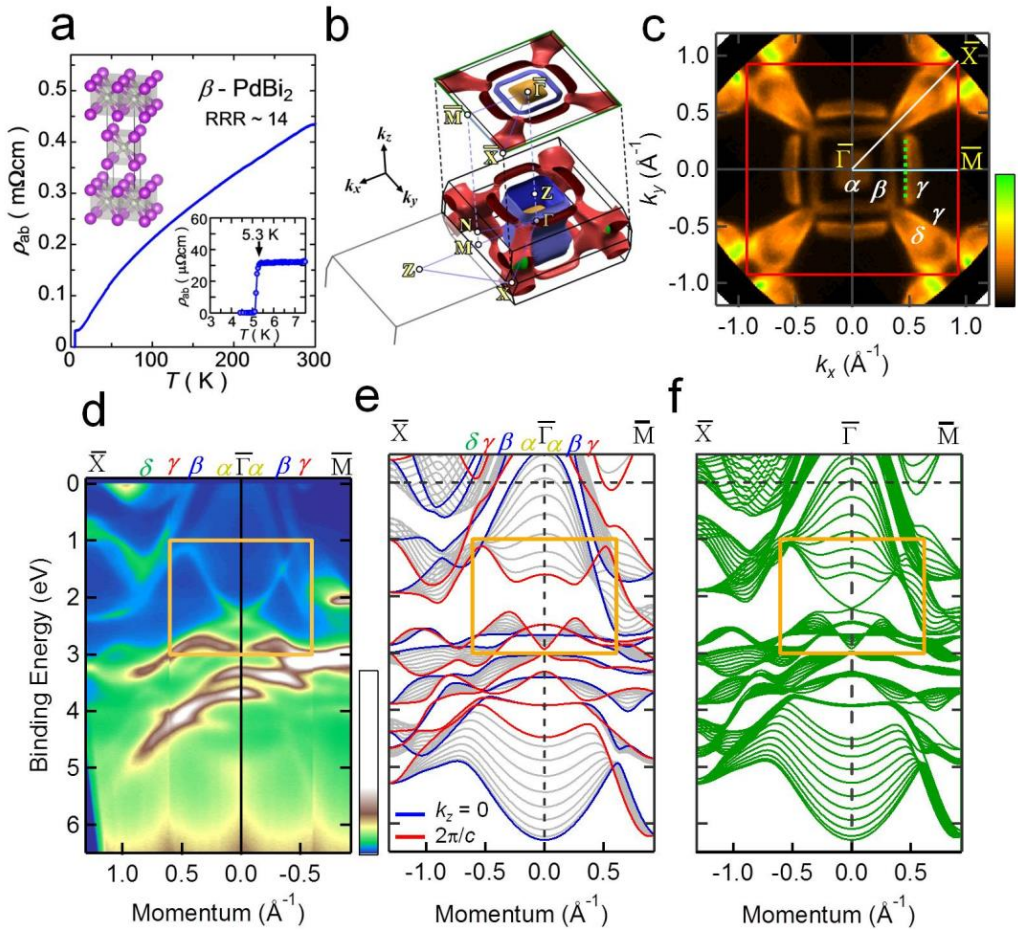

**Figure 1 | Electronic structure of superconductor $\beta$-PdBi$_2$**

**a**, In-plane electrical resistivity ($\rho_{ab}$) indicated with the crystal structure (left). The right inset shows $\rho_{ab}$ near $T_c$ = 5.3 K. **b**, Calculated Fermi surfaces shown with the first BZ. Square plane with $\overline{\Gamma}$, $\overline{M}$ and $\overline{X}$ represents the 2D-projected surface BZ. **c**, 4-fold symmetrized FS obtained by mapping ARPES intensity at $E_F \pm 8$ meV. Two electron-like and two hole-like FS are denoted by $\alpha$, $\beta$ and $\gamma$, $\delta$, respectively. **d**, ARPES image recorded along $\overline{X}$-$\overline{\Gamma}$ (left) and $\overline{\Gamma}$-$\overline{M}$ (right) cuts, shown as the white and light blue lines in **c**. **e**, Calculated bulk band dispersions projected onto 2D surface BZ. Blue (Red) curves correspond to $k_z = 0$ ($2\pi/c$). **f**, Surface band dispersions obtained by slab calculation of 11 PdBi$_2$-layers. Orange rectangles in **d-f** indicate the region where S-DC appears.

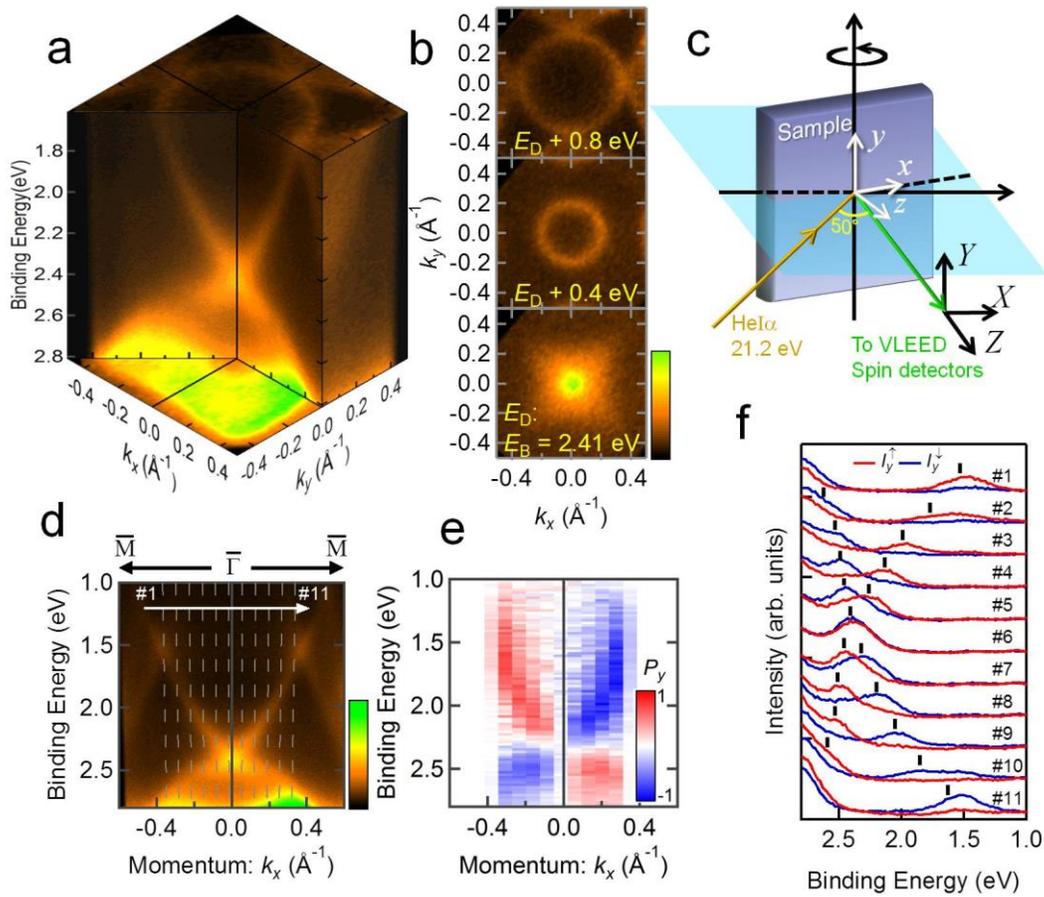

**Figure 2 | Band dispersion and spin polarization of surface Dirac-Cone band**

**a**, Close-up of the observed surface Dirac-cone (S-DC) dispersions. **b**, Constant energy cuts of S-DC at $E_B$ = $E_D$ +0.8 eV, $E_D$ + 0.4 eV and $E_D$ (= 2.41 eV), respectively. **c**, Schematic of SARPES experimental geometry. **d**, ARPES image of S-DC along $\overline{\Gamma}$-$\overline{M}$. Gray lines (#1 ~ #11) represent the measurement cuts for SARPES EDCs shown in **f**. **e**, SARPES image of S-DC dispersions for spin $y$-component. Red-blue color-scale indicates the polarization. **f**, SARPES EDCs for momenta #1 ~ #11 in **d**, respectively. Red (blue) curves show the spin-up (spin-down) components. The black markers show the peak positions of the ARPES intensity in **d**.

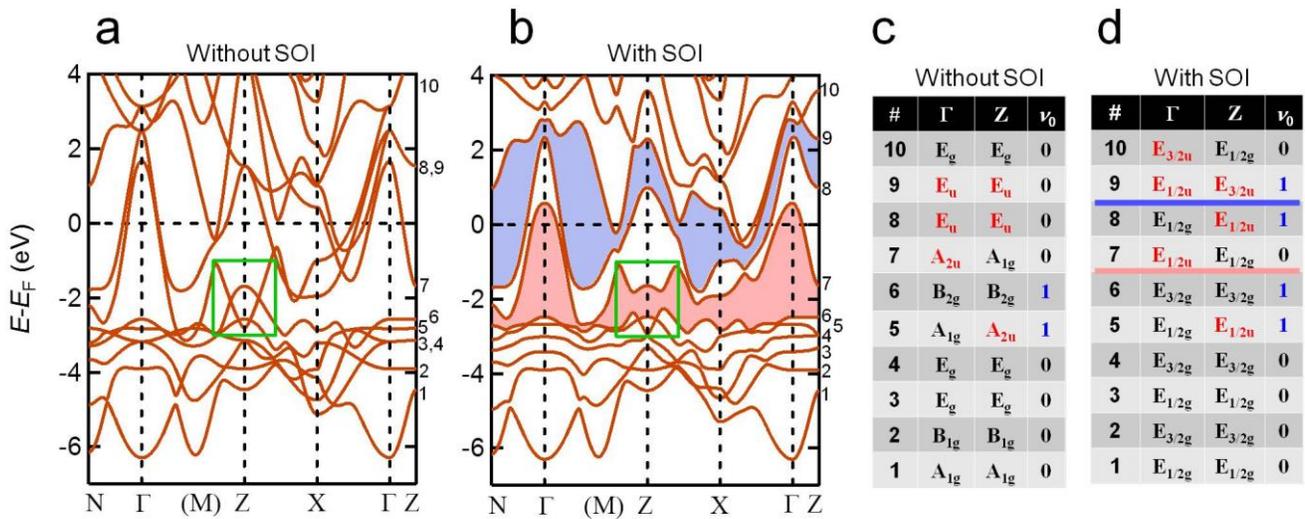

**Figure 3 | Parity and $\nu_0$ analyses for valence bands without and with SOI**

**a**,**b**, First-principles band calculations without and with SOI, respectively. Valence bands are numbered by the energy at Z-point, as shown in the right side of panels. Red (blue) shaded area in **b** shows "gap7-6" ("gap9-8") induced by SOI. **c**,**d**, Tables of $\nu_0$ and the symmetries of wave-functions at Γ- and Z-points without and with SOI, respectively. Indicated with red has odd-parity. $\nu_0 = 1$ indicates the topologically nontrivial band-inverted state.

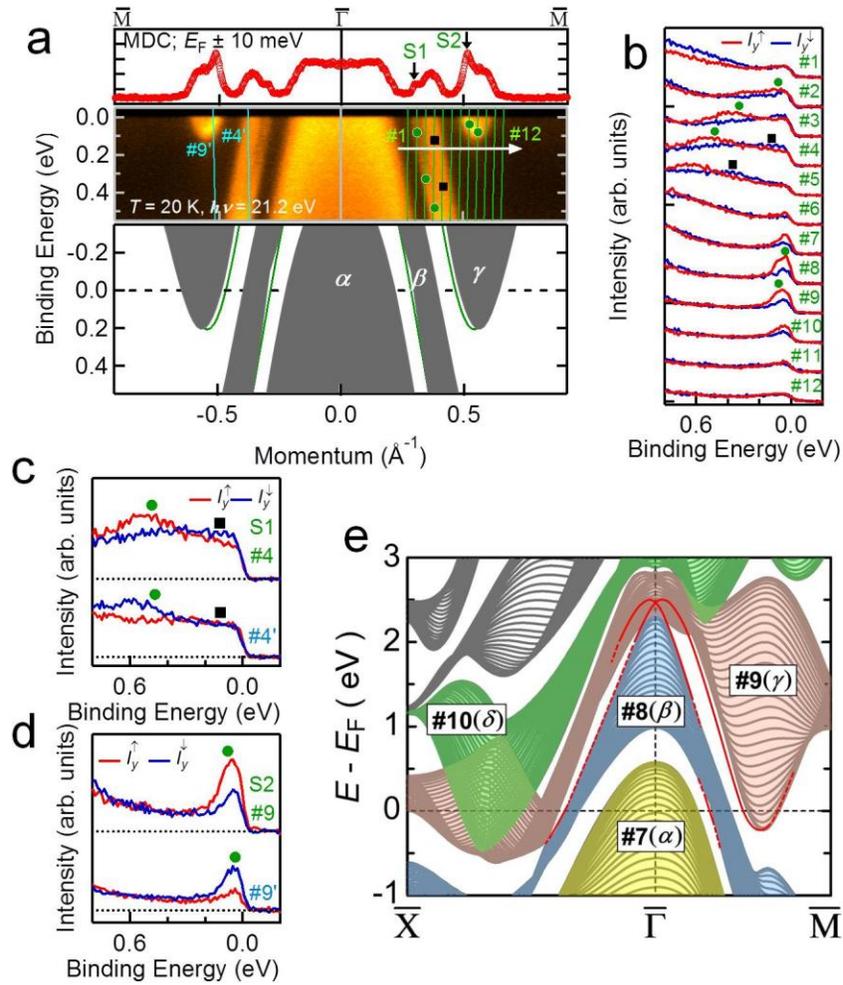

**Figure 4 | Spin-polarized surface states crossing $E_F$**

**a**, MDC at $E_F$ (top), ARPES image (middle), and calculation (bottom) are shown along $\overline{\Gamma}$-$\overline{M}$. The black arrows of MDC indicate the intensities from surface bands denoted by S1 and S2. Green circles (black squares) depicted on the ARPES image are peak positions of EDCs and MDCs for surface (bulk) bands. In the calculation, surface band dispersions (green) are overlaid to 2D-projected bulk bands (gray). **b**, Spin-resolved EDCs for spin *y*-component recorded at momenta #1 ~ #12 in **a**, respectively. **c**, Spin-resolved EDCs for S1 at momenta #4 and #4' in **a**. **d**, Spin-resolved EDCs for S2 at momenta #9 and #9'. Peak positions of ARPES intensity in **b**-**d** are depicted similarly to **a**. **e**, 2D-projected bulk bands with the surface states crossing $E_F$ (highlighted by red), namely the trivial S1 and the topological S2 connecting B9 and B8, from the slab calculations.